\documentclass[conference]{IEEEtran}
\IEEEoverridecommandlockouts
\usepackage{cite}
\usepackage{amsmath,amssymb,amsfonts}
\usepackage{algorithmic}
\usepackage{algorithm}
\usepackage{graphicx}
\usepackage{textcomp}
\usepackage{xcolor}
\usepackage{booktabs}
\usepackage{multirow}
\usepackage{hyperref}
\usepackage[caption=false,font=normalsize,labelfont=sf,textfont=sf]{subfig}
\usepackage{stfloats}
\usepackage{url}
\usepackage{verbatim}
\usepackage{marvosym}
\usepackage{makecell}
\usepackage{svg}

\usepackage{tabularx}
\usepackage{caption}
\usepackage{tikz}
\usepackage{threeparttable}
\makeatletter
\def\ps@IEEEtitlepagestyle{%
  \def\@oddfoot{}% 清空页脚
  \def\@evenfoot{}%
  \def\@oddhead{\hfill\@IEEEpubid\hfill}% 将pubid移至页眉
  \def\@evenhead{\hfill\@IEEEpubid\hfill}%
}
\makeatother

\IEEEpubid{\footnotesize\textit{This work has been submitted to the IEEE for possible publication.\\
Copyright may be transferred without notice, after which this version may no longer be accessible.}}
\def\BibTeX{{\rm B\kern-.05em{\sc i\kern-.025em b}\kern-.08em
    T\kern-.1667em\lower.7ex\hbox{E}\kern-.125emX}}
\begin{document}

\title{ReasoningV: Efficient Verilog Code Generation with Adaptive Hybrid Reasoning Model
}
\definecolor{LJJ}{RGB}{255,0,0}
\definecolor{XZW}{RGB}{0,255,0}

\author{
\IEEEauthorblockN{
Haiyan Qin\IEEEauthorrefmark{1},
Zhiwei Xie\IEEEauthorrefmark{1},
Jingjing Li\IEEEauthorrefmark{1},
LiangChen Li\IEEEauthorrefmark{1},
Xiaotong Feng\IEEEauthorrefmark{1},
Junzhan Liu\IEEEauthorrefmark{1},
Wang Kang\IEEEauthorrefmark{1}\IEEEauthorrefmark{4}
}
\IEEEauthorblockA{
\IEEEauthorrefmark{1}
\textit{National Key Laboratory of Spintronics, Hangzhou International Innovation Institute;}\\
\textit{School of Integrated Circuit Science and Engineering, Beihang University, China}
}
\IEEEauthorblockA{
\IEEEauthorrefmark{1}
\{\textit{haiyanq, xiezhiwei, LIJJ57,liliangchen, betty513, liujunzhan, wang.kang}\}@buaa.edu.cn
}
\thanks{\IEEEauthorrefmark{4} Denotes Corresponding Authors.}
\vspace{-3em}
}

\maketitle

\begin{abstract}
Large Language Models (LLMs) have advanced Verilog code generation significantly, yet face challenges in data quality, reasoning capabilities, and computational efficiency. This paper presents ReasoningV, a novel model employing a hybrid reasoning strategy that integrates trained intrinsic capabilities with dynamic inference adaptation for Verilog code generation. Our framework introduces three complementary innovations: (1) ReasoningV-5K, a high-quality dataset of 5,334 functionally verified instances with reasoning paths created through multi-dimensional filtering of PyraNet samples; (2) a two-stage training approach combining parameter-efficient fine-tuning for foundational knowledge with full-parameter optimization for enhanced reasoning; and (3) an adaptive reasoning mechanism that dynamically adjusts reasoning depth based on problem complexity, reducing token consumption by up to 75\% while preserving performance. Experimental results demonstrate ReasoningV's effectiveness with a pass@1 accuracy of 57.8\% on VerilogEval-human, achieving performance competitive with leading commercial models like Gemini-2.0-flash (59.5\%) and exceeding the previous best open-source model by 10.4 percentage points. ReasoningV offers a more reliable and accessible pathway for advancing AI-driven hardware design automation, with our model, data, and code available at \url{https://github.com/BUAA-CLab/ReasoningV}.
\end{abstract}

\begin{IEEEkeywords}
Large Language Models, Verilog Code Generation, Verifiable Dataset, Two-Stage Training, Adaptive Reasoning
\end{IEEEkeywords}

\section{Introduction}

Hardware Description Languages (HDLs) like Verilog are crucial for digital circuit design, but manual creation is increasingly challenged by circuit complexity, leading to time-consuming, error-prone processes and prolonged verification cycles \cite{chip, aivril, openabc, verigen}. While Electronic Design Automation (EDA) tools handle low-level optimization, generating Register-Transfer Level (RTL) code from natural language specifications remains a bottleneck \cite{Openllm, verilogcoder, EDA}. Recent LLMs show promise in automating this translation \cite{evaluating}\cite{autovcoder}, with some achieving notable successes like processor tape-outs \cite{chip} or integrating verification \cite{aivril}\cite{ verilogcoder}. However, significant hurdles persist. Critically, the field suffers from \textbf{data quality deficiency}, relying on datasets often containing unverified or flawed samples that hinder learning complex hardware logic \cite{Openllm}\cite{rtlcoder}. Compounding this, current models possess \textbf{limited reasoning capabilities} for intricate hardware tasks like multi-module integration or complex state machines, impacting functional correctness \cite{verilogeval, rtllm, rtl-repo}. Finally, \textbf{computational inefficiency} arises from resource-intensive inference processes and inflexible, uniform reasoning depths \cite{rtlcoder}\cite{ betterv}.

To overcome these limitations, we propose \textit{ReasoningV}, the first \textbf{hybrid reasoning model} for Verilog code generation. It uniquely combines intrinsic reasoning capabilities developed through training with \textbf{adaptive inference strategies that dynamically adjust or bypass the reasoning process based on problem complexity}, optimizing both performance and efficiency. ReasoningV introduces three core innovations:
\begin{enumerate}
    \item \textbf{ReasoningV-5K Dataset}: A high-quality, functionally verified dataset of 5,334 problem-solution-test triplets derived from 690K samples, tackling the data quality deficiency.
    \item \textbf{Two-Stage Training}: A method combining parameter-efficient fine-tuning for foundational knowledge with full-parameter optimization for deep reasoning, addressing limited reasoning capabilities.
    \item \textbf{Adaptive Reasoning Mechanism}: A dynamic system adjusting reasoning depth based on complexity, reducing token consumption by up to 78\% while maintaining quality, resolving computational inefficiency.
\end{enumerate}

Experiments show ReasoningV achieves state-of-the-art (SOTA) results among open-source models, notably reaching 57.8\% pass@1 on VerilogEval-Human\cite{verilogeval}, surpassing the previous best by 10.4 percentage points \cite{origen}. This work demonstrates that targeted methodological ingenuity in data curation, training, and inference can markedly improve the fidelity and practicality of AI-generated hardware descriptions, advancing reliable and resource-efficient AI tools for the hardware development lifecycle. The remainder of this paper details our  related work, methodology, and results.

\section{Related Work}

\subsection{LLMs and Verilog Generation Challenges}

\begin{table}[htbp]
    \centering
    \caption{Verilog Datasets for LLMs}
    \label{tab:dataset_comparison}
    \renewcommand{\arraystretch}{1.0}
    \setlength{\tabcolsep}{4pt}
    \footnotesize
    \begin{tabular*}{\hsize}{@{}@{\extracolsep{\fill}}lllllllllllll@{}}
    \toprule
    \textbf{Dataset} & \textbf{Source} & \textbf{Size} & \textbf{Key Features} \\
    \midrule
    PyraNet\cite{PyraNet} & GitHub+LLM & 690K+ & Syntax-checked, quality-ranked \\
    MG-Verilog\cite{mg} & Open-Source & 11K+ & Multi-grain descriptions \\
    RTLCoder\cite{rtlcoder} & LLM-gen & 80K+ & ABV/FPV subset available \\
    CodeV\cite{codev} & GitHub & 165K & Multi-level summaries \\
    VerilogEval\cite{verilogeval} & HDLBits & 330 & Testbench evaluation \\
    \midrule
    \textbf{ReasoningV} & PyraNet & 5K & \textbf{Problem-solution-test triples} \\
    \bottomrule
    \end{tabular*}
    \vspace{0.5em}
    \scriptsize
    ABV: Assertion-Based Verification; FPV: Formal Property Verification
\end{table}

While LLMs excel at general code generation, producing syntactically correct and functionally robust code across high-level languages like Python and Java \cite{Codex}, their application to HDLs like Verilog presents unique challenges. These stem from hardware's inherent parallelism, strict timing constraints, and the need for precise synchronization in concurrent operations, which differ significantly from the sequential logic of software programming \cite{ML4EDA}\cite{ revisiting}. Verilog’s syntax and semantics require models to capture hardware-specific behaviors, such as clock-driven state transitions and resource constraints, which are poorly represented in general-purpose code datasets \cite{Openllm}. Furthermore, the relative scarcity of high-quality, functionally verified training data exacerbates these issues, as existing datasets often lack rigorous verification, hindering the learning of correct hardware logic \cite{ML4EDA}.

% Another critical challenge is the complexity of translating natural language specifications into Verilog code that meets functional and performance requirements. Unlike software, where functionality can often be validated through unit tests, hardware designs demand comprehensive verification to ensure correctness under diverse operating conditions, such as varying clock frequencies or input patterns \cite{revisiting}. The lack of standardized, high-quality datasets tailored to Verilog’s hardware-specific needs limits LLMs’ ability to generalize across diverse design types, from simple combinational circuits to complex multi-module systems with asynchronous clock domains \cite{verilogeval}. Additionally, LLMs often struggle with reasoning about hardware-specific optimizations, such as minimizing latency or power consumption, which are critical in real-world hardware design but underrepresented in training data \cite{betterv}. These challenges highlight the need for specialized approaches to enhance LLMs’ capabilities in Verilog generation, addressing both data quality and the intrinsic complexities of hardware design.

\subsection{Verilog Datasets and the Need for Quality}
\label{sec:related_datasets}

% --- Keep Table III ---
% --- Comparative Analysis Table ---
% --- Comparative Analysis Table (Condensed Version) ---
\begin{table*}[htbp]
    \centering
    \caption{Comparative Analysis of Recent Verilog Generation Systems} 
    \label{tab:system_comparison} 
    \small
    \setlength{\tabcolsep}{3pt}
    \begin{tabular*}{\hsize}{@{}@{\extracolsep{\fill}}lllllllllllll@{}}
    \toprule
    \textbf{System} & 
    \textbf{Base LLM} & 
    \textbf{Key Innovation(s)} & 
    \textbf{Training Method} & 
    \textbf{Reasoning Approach}  \\
    \midrule
    \textbf{ReasoningV} (Ours) & 
    Qwen2.5-Coder-7B & 
    ReasoningV-5K, Adaptive Infer & 
    2-Stage (LoRA+Full) & 
    Trained + Adaptive  \\ 
    
    HDLCoRe \cite{HDLCoRe} & 
    LLM-agnostic & 
    HDL-CoT, Self-Verify, RAG & 
    Training-Free & 
    CoT + Self-Verify \\ 
    
    VeriMind \cite{VeriMind} & 
    LLM-agnostic &
    Agentic Framework, ToT-like Reason & 
    Training-Free & 
    Agent-driven Multi-step  \\ 
    
    Paradigm-Based\cite{Paradigm} & 
    LLM-agnostic &
    Paradigm Blocks, 2-Phase Workflow & 
    Training-Free & 
    Prompt (Paradigm Blocks)  \\ 
    
    CodeV \cite{codev} & 
    Various (e.g., DeepSeek) & 
    Multi-level Summarization & 
    Instruction Tuning & 
    Learned (Instruction Tune) \\ 
    
    VeriSeek \cite{VeriSeek} & 
    DeepSeekCoder-6.7B & 
    RL Post-Train, AST Reward & 
    Pre-train+Tune+RL & 
    Learned (RL)  \\ 
    
    RTLCoder \cite{rtlcoder} & 
    DeepSeek/Mistral-7B & 
    Data Aug, Quality Feedback & 
    Instruct Tune & 
    Learned (Instruct Tune) \\
    \bottomrule
    \end{tabular*}
    \begin{tablenotes}
    \small
    \item \textbf{Abbr.:} RAG = Retrieval-Augmented Generation ToT = Tree-of-Thought
    \item RL = Reinforcement Learning, AST = Abstract Syntax Tree, Infer = Inference, Aug = Augmentation
    \end{tablenotes}
\end{table*}

Training data quality is crucial for Verilog generation \cite{Kodcode}. As summarized in Table~\ref{tab:dataset_comparison}, current datasets like PyraNet\cite{PyraNet}, MG-Verilog\cite{ mg}, and RTLCoder-Data\cite{ rtlcoder} primarily rely on syntax checks or LLM-based ranking, often lacking functional verification and containing redundancy . Others like CodeV\cite{codev} infer quality from source code . While evaluation benchmarks like VerilogEval\cite{verilogeval} provide verified problems, they are insufficient for training complex reasoning. This highlights a gap: the need for a dataset combining guaranteed functional correctness with structures for learning hardware reasoning.

ReasoningV-5K addresses this gap. Derived from filtering PyraNet\cite{PyraNet}, it provides 5,334 high-quality ``problem-solution-test'' triplets. Crucially, \textbf{every solution is functionally verified via simulation} using its corresponding testbench, and each sample includes a \textbf{distilled reasoning path}. This combination directly tackles data quality, verification, and reasoning deficits needed for more reliable Verilog generation models.

\subsection{Verilog Generation Models and Reasoning Limitations}

Data-driven Verilog generation aims to overcome traditional EDA limitations \cite{synopsys}. While recent models (e.g., RTLCoder \cite{rtlcoder}, VeriSeek \cite{VeriSeek}) show progress, many still struggle with \textbf{limited reasoning capabilities} for complex hardware tasks, impacting functional correctness \cite{verilogeval, rtllm}. Recent advancements (Table~\ref{tab:system_comparison}) explore \textit{training-free} methods using prompting or agents (e.g., HDLCoRe \cite{HDLCoRe}, VeriMind \cite{VeriMind}) and \textit{training-based} approaches like instruction tuning or RL (e.g., CodeV \cite{codev}, VeriSeek \cite{VeriSeek}).

ReasoningV adopts a distinct \textbf{hybrid reasoning} strategy. Unlike purely prompt-driven methods, it invests in \textit{training} intrinsic reasoning using the \textbf{ReasoningV-5K dataset} and a \textbf{two-stage training method}. It aims for greater robustness than prompt-elicited reasoning and more targeted enhancement than standard tuning or RL focused mainly on code structure or basic correctness.

\subsection{Adaptive Reasoning for Efficiency}
\label{sec:related_reasoning_enhancement} % Renamed label slightly for uniqueness

Enhancing LLM reasoning often incurs significant \textbf{computational inefficiency} if applied uniformly \cite{Open-O1}, especially via methods like Chain-of-Thought (CoT) \cite{HDLCoRe}. Efforts to improve efficiency include advanced prompting, architectural/training innovations, and adaptive inference techniques like dynamic routing or complexity-based resource allocation \cite{VeriSeek,stop,qu2025survey,dynamic}.

Building on these, ReasoningV introduces an \textbf{adaptive reasoning mechanism} (Sec~\ref{sec:adaptive_reasoning}). It uses a lightweight classifier (Judge Adapter) to estimate problem complexity (Easy/Medium/Hard) and dynamically allocates reasoning budget/mode. This approach, related to complexity-based routing \cite{dynamic} but integrated with our trained reasoning capabilities, optimizes resource use across different problem difficulties in Verilog generation, achieving significant token savings (up to 78\%) while maintaining performance.
\begin{figure*}[htbp]
    \centering
    \includegraphics[width=\textwidth]{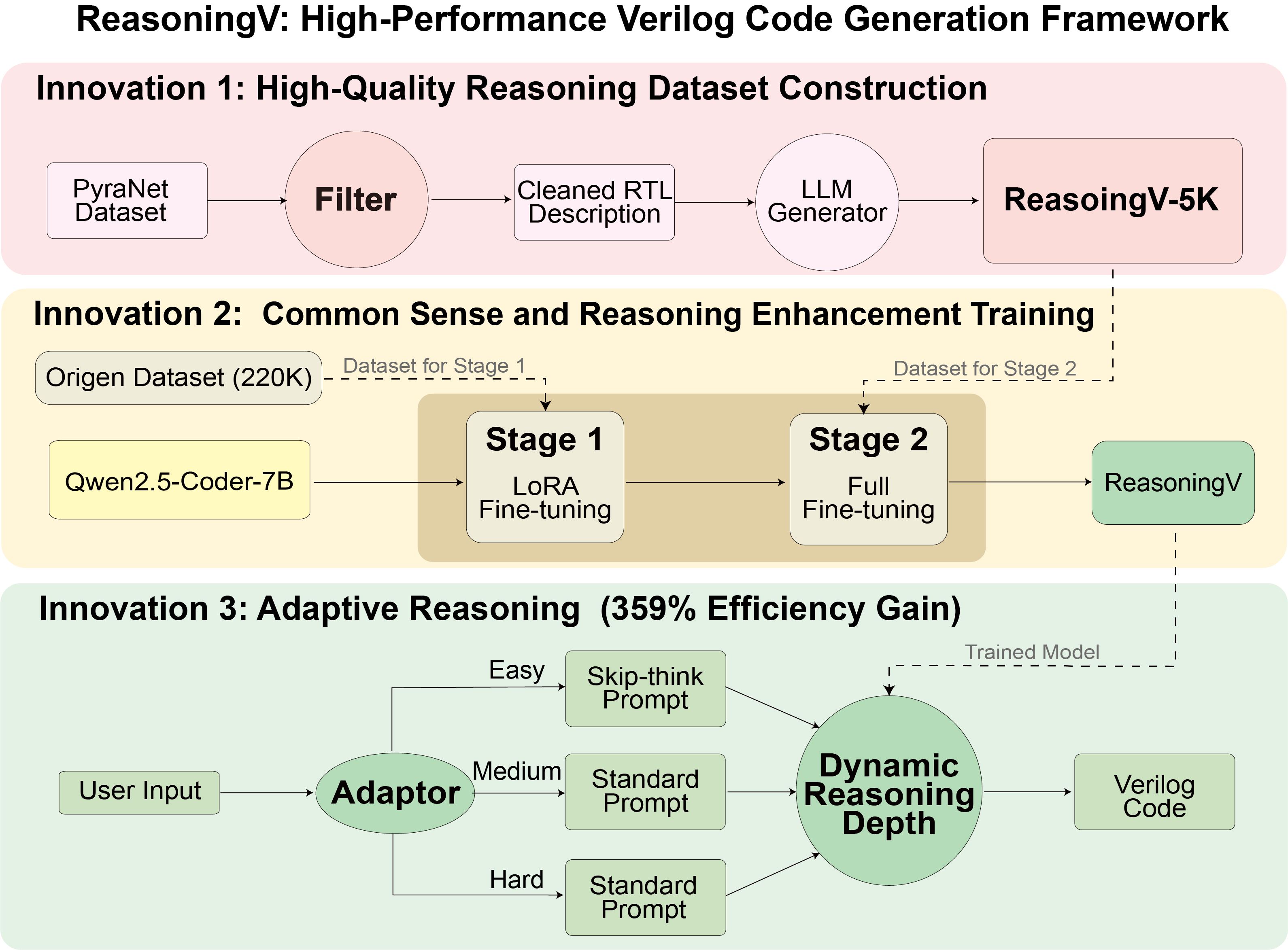}
    \caption{Overall architecture of the ReasoningV framework.}
    \label{fig:framework}
\vspace{-1em}
\end{figure*}
\section{Methodology}
This section outlines the ReasoningV framework for Verilog code generation. As shown in Fig.\ref{fig:framework}, our approach comprises three core components: constructing the ReasoningV-5K dataset, implementing a two-stage training method, and developing an adaptive reasoning mechanism. These elements systematically address poor data quality, limited reasoning capabilities, and computational inefficiency through data refinement, staged training, and dynamic reasoning optimization.

\subsection{High-Quality Reasoning Dataset Construction}
\label{sec:dataset_construction}

\begin{figure*}[htbp]
   \centering
   \includegraphics[width=\textwidth]{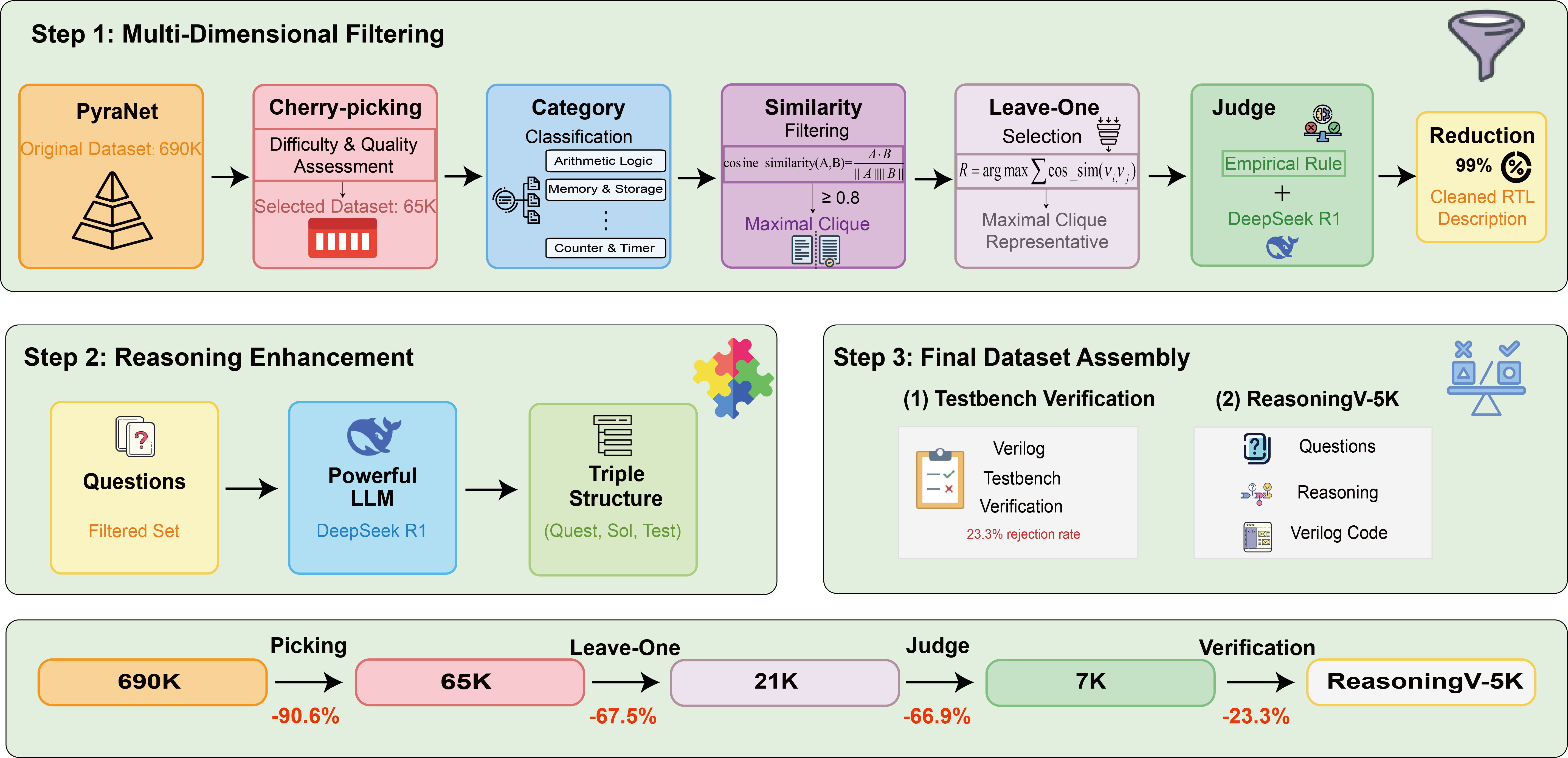}
   \caption{Multi-stage data filtering pipeline for ReasoningV-5K dataset construction.}
   \label{fig:dataset_construction}
\end{figure*}

High-quality training data is vital for effective HDL generation. We constructed ReasoningV-5K, a rigorously verified Verilog reasoning dataset containing 5,334 high-quality samples, through multi-stage filtering and verification processes designed to enhance data reliability and enforce design best practices. As illustrated in Fig.~\ref{fig:dataset_construction}, our construction process consists of three main steps: (1) Multi-Dimensional Filtering of PyraNet samples\cite{PyraNet}, (2) Reasoning Enhancement with DeepSeek-R1\cite{deepseekr1}, and (3) Final Dataset Assembly with functional testbench verification.

\subsubsection{Multi-Dimensional Filtering Process}

\textbf{PyraNet Dataset Selection and Initial Filtering}\\
We started with the PyraNet dataset \cite{PyraNet} (approx. 690K samples), acknowledging its scale but also its limitations (redundancy, quality issues). Filtering was essential for quality and cost-effectiveness, aligning with findings that quality outweighs quantity in HDL training \cite{S1, rtllm}. We focused on the Tier2 subset, retaining only compilable samples (65,344 candidates).

\textbf{Domain-Specific Redundancy Elimination}\\
We classified candidates into 15 hardware domains (using Qwen2.5-32B) and applied a three-step redundancy elimination within each domain: (1) Cosine similarity calculation (threshold 0.8, empirically determined) for code and descriptions; (2) Maximal clique grouping using Bron-Kerbosch; (3) Selecting one representative per clique using a ``leave-on'' rule. This yielded approx. 21K unique samples.

\textbf{Expert-Guided Automated Quality Assessment}\\
We further refined the dataset to 12K samples using DeepSeek-R1, prompted with expert-defined rules covering naming conventions and RTL coding style best practices (e.g., module instantiation, state machine structure). Only samples conforming to these guidelines were retained, ensuring high code quality and readability.

\subsubsection{Compression Ratio Analysis}  
We validated our filtering effectiveness using Compression Ratio (CR), which measures the ratio of original text size to gzip-compressed text size, reflecting information density and non-redundancy:

\begin{table}[htbp]
   \centering
   \caption{Compression Ratio comparison across Verilog datasets}
   \label{tab:compression_ratio}
   \begin{tabular}{@{}lcc@{}}
   \toprule
   \textbf{Dataset} & \textbf{CR} & \textbf{CR-POS} \\
   \midrule
   RTLCoder-27K & 4.41 & 7.61 \\
   Goh et al. & 5.27 & 10.1 \\
   MG-Verilog & 5.80 & 9.16 \\
   Magicodes-OSS-Instance & 4.02 & 6.67 \\
   ReasoningV-5K (Filtered) & 4.56 & 6.51 \\
   \bottomrule
   \end{tabular}
\end{table}

ReasoningV-5K's CR of 4.56, compared to RTLCoder-27K (4.41) and MG-Verilog (5.80), indicates high information density with reduced redundancy (Table \ref{tab:compression_ratio}). This analysis confirms that our filtering process effectively maintained high information value while removing repetitive content, resulting in a dataset with balanced information density that avoids both excessive redundancy and information loss.

\subsubsection{Dataset Verification and Structure}
\begin{figure}[htbp]
   \centering
   \includegraphics[width=\columnwidth]{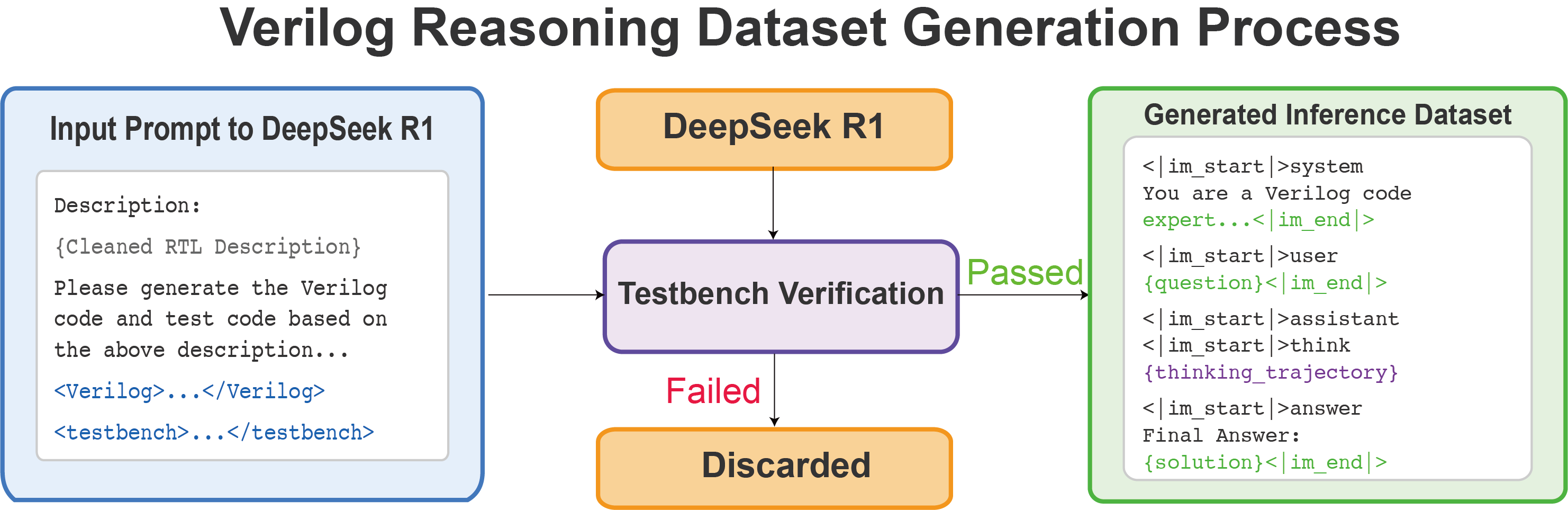}
   \caption{Verilog Reasoning Path and Testbench Generation.}
   \label{fig:data_generation}
\end{figure}

% problem-description-reasoning\_path-solution-testbench
We adopted a ``problem-description-reasoning\_path-solution-testbench'' format. Each entry includes the problem description, a detailed reasoning path (distilled using DeepSeek-R1, see Fig~\ref{fig:data_generation}), the Verilog solution, and a corresponding functional testbench (also generated with DeepSeek-R1 guidance).

Crucially, \textbf{functional verification} was performed by simulating each solution against its testbench using Icarus Verilog \cite{icarus}. Samples failing simulation (approx. 23.3\% of the 7K candidates) were discarded. This prioritization of functional accuracy resulted in the final ReasoningV-5K dataset: 5,334 high-quality, functionally verified instances with reasoning paths, exhibiting high information density (CR 4.56).

 Our filtering reduced the initial 690K samples by approx. 99.3\%, significantly improving quality and ensuring functional correctness per sample.

\subsection{Two-Stage Training Method}
\label{sec:two_stage_training}

ReasoningV employs a two-stage training strategy targeting distinct goals: broad domain knowledge acquisition (Stage 1) and deep reasoning capability development (Stage 2). This staged approach optimizes both effectiveness and efficiency.

\subsubsection{Motivation and Efficiency Analysis}
Stage 1 efficiently establishes foundational knowledge (Verilog syntax, common patterns) using parameter-efficient tuning, while Stage 2 focuses on complex reasoning (e.g., multi-module integration) using full-parameter optimization on high-quality data. Stage 1 trains only 1.05\% of parameters, contrasting with Stage 2's full optimization.

\subsubsection{Training Methodology}
Distinct configurations were used for each stage, detailed in Table~\ref{tab:training_details}.

\begin{table}[htbp]
    \centering
    \caption{Comparison of two-stage training configurations}
    \label{tab:training_details}
    \small
    \begin{tabular}{@{}p{2.9cm}p{2.4cm}p{2.4cm}@{}}
    \toprule
    \textbf{Configuration} & \textbf{Stage 1} & \textbf{Stage 2} \\
    \midrule
    Training Approach & LoRA & Full-Parameter \\
    Trainable Parameters & 80.7M (1.05\%) & 7.7B (100\%) \\
    Dataset & OriGen (220K) & ReasoningV-5K \\
    Sequence Length & 2,048 tokens & 8,192 tokens \\
    Micro-Batch Size & 2 & 1 \\
    GA & 16 & 32 \\
    Effective Batch Size & 32 & 64 \\
    Learning Rate & 1e-5 & 1e-5 \\
    LR Scheduler & Cosine Decay & Cosine Decay \\
    Warmup Ratio & 0.03 & 0.05 \\
    Adam $\beta_1$/$\beta_2$ & 0.9/0.999 & 0.9/0.95 \\
    Training Epochs & 5 & 5 \\
    GPUs & 1 A100-80GB & 2 A100-80GB \\
    Precision & BFloat16 & BFloat16 \\
    Distributed Training & None & FSDP \\
    Loss Function & Cross-Entropy & Cross-Entropy \\
    \bottomrule
    \end{tabular}
    \begin{tablenotes}
    \small
    \item Abbr.: GA = Gradient Accumulation
    % \item Abbr.: RV = ReasoningV, S1 = Stage1, RO = Reasoning-Only. Bold values indicate best results within each category.
    \end{tablenotes}
\end{table}

\subsubsection{Commonsense Training Stage}
Stage 1 builds foundational Verilog knowledge on the Qwen2.5-Coder-7B base model \cite{qwen2}. We applied Low-Rank Adaptation (LoRA) \cite{lora} for parameter-efficient fine-tuning (PEFT), targeting key projection layers (details omitted for brevity, see \cite{lora}). With rank=32 and $\alpha=32$, this resulted in only 80.7M trainable parameters (1.05\% of total), enabling effective adaptation with minimal cost using the OriGen dataset. The training objective was the standard causal language modeling loss (Cross-Entropy), maximizing the likelihood of target tokens while masking input tokens during loss calculation.

\subsubsection{Reasoning Enhancement Training Stage}
Stage 2 enhances reasoning via full-parameter fine-tuning on the ReasoningV-5K dataset. We first merged the learned LoRA adapter weights into the base model weights ($W_{\text{stage2\_init}} = W_0 + BA$). Then, all 7.7B parameters were trained to deeply integrate complex reasoning patterns from ReasoningV-5K. The same Cross-Entropy loss objective was used, now focusing on generating reasoning paths and correct code. To manage full-parameter training efficiently on the 5K dataset, we employed Fully Sharded Data Parallel (FSDP) \cite{pytorchfsdp2022} for distributed training across GPUs, BFloat16 precision for speed and memory efficiency on A100 GPUs, and the AdamW optimizer \cite{adamw} (specific betas in Table~\ref{tab:training_details}). These techniques ensured effective learning while maintaining generalization.

\subsection{Adaptive Reasoning Method}
\label{sec:adaptive_reasoning}

To efficiently evaluate problem complexity and tailor the generation process, we introduce a lightweight Judge LoRA adapter mounted on the ReasoningV backbone model. This parameter-efficient tuning strategy maintains the primary model parameters unchanged, effectively mitigating catastrophic forgetting while augmenting the model's capability for problem difficulty classification.

\subsubsection{Judge Adapter Training}
We systematically sampled 10,000 representative questions from the training corpus and constructed a specialized difficulty-classification dataset through knowledge distillation from the DeepSeek-V3 model. This engineered dataset was then employed to perform LoRA training on our base model, yielding a Judge Adapter specifically optimized for hierarchical question difficulty assessment. The adapter categorizes each input into one of three difficulty levels: “Easy”, “Medium”, or “Hard”, enabling fine-grained control over downstream reasoning strategies.

To better align inference strategies with difficulty levels, we introduced dynamic mode selection based on the Judge Adapter’s assessment:
\begin{itemize}
    \item \textbf{Easy/Direct Mode (``Easy''):} Bypasses explicit reasoning. Minimal prompt. Focus on code output. Lower token budget (\texttt{max\_new\_tokens=512}).
    \item \textbf{Medium/Standard Reasoning Mode (``Medium''):} Standard reasoning prompt. Balanced token budget (\texttt{max\_new\_tokens=1280}) for reasoning and code.
    \item \textbf{Hard/Extended Reasoning Mode (``Hard''):} Detailed decomposition. Increased token budget (\texttt{max\_new\_tokens=4096}) for extensive reasoning and complex code.
\end{itemize}

\subsubsection{Reasoning Mode Selection and Application Logic}
As illustrated in Fig.\ref{fig:framework} (Innovation 3), the framework uses the Judge Adapter's output to dynamically select reasoning modes and allocate resources:

\textbf{Operational Phases and Task Division}
The system operates in two modes:
\begin{itemize}
    \item \textbf{Code Generation:} Disable Judge Adapter. Use full ReasoningV model.
    \begin{itemize}
        \item Input: Problem description (+ optional reasoning prompt).
        \item Output: Reasoning path (if applicable) and Verilog code.
        \item Purpose: Generate solution based on selected strategy.
    \end{itemize}
\end{itemize}

\textbf{Resource Optimization}
This adaptive approach significantly optimizes resource use:
\begin{itemize}
    \item The lightweight Judge Adapter is active only briefly
    \item Conditional generation allocates compute efficiently:
    \begin{itemize}
        \item More resources for hard problems
        \item Fewer resources for easy ones
    \end{itemize}
    \item Optimizes throughput and cost compared to fixed-depth reasoning approaches
\end{itemize}

\section{Experiments}

In this section, we evaluate ReasoningV's performance against baseline models and conduct detailed ablation studies to analyze the impact of different components in our framework. We first describe our experimental setup, including benchmarks and implementation details. Then, we present comparative results against SOTA models and analyze ReasoningV's components through systematic ablation studies.

\subsection{Experimental Setup}

\subsubsection{Benchmarks}

We evaluate ReasoningV on three widely-used benchmarks for Verilog code generation:

\begin{itemize}
    \item \textbf{VerilogEval-human}~\cite{verilogeval}: This benchmark consists of 156 Verilog problems created by hardware engineers, designed to evaluate models on real-world hardware design tasks.
    
    \item \textbf{VerilogEval-machine}~\cite{verilogeval}: This benchmark includes 143 synthetic Verilog problems generated from templates, focused on evaluating models on standard hardware design patterns.
    
    \item \textbf{RTLLM}~\cite{rtllm}: A comprehensive benchmark containing 30 real-world Verilog problems covering a diverse range of hardware design tasks with varying complexity levels.
\end{itemize}

Following standard practices in the field~\cite{ origen}\cite{ betterv}, we use the widely-adopted Pass@k metric in code generation tasks:

\begin{equation}
\text{pass@k} := \mathbb{E}_{\text{Problems}}\left[1 - \frac{\binom{n-c}{k}}{\binom{n}{k}}\right]
\end{equation}
where k is the number of samples to generate for each problem, and $p_i$ represents the probability that a randomly sampled solution is correct. In our experiment, we set n=10 total samples per task and estimate $p_i$ as the fraction of correct solutions among these samples. The Pass@k metric reflects the expected probability of solving a task if we sample k times and take the best solution.

\subsubsection{Implementation Details}

We implemented ReasoningV based on the Qwen2.5-Coder-7B model~\cite{qwen2} using the PyTorch framework. For inference, we used standard parameters optimized for code generation: temperature = 0.2, top\_p = 0.95. All experiments were conducted on NVIDIA A100-80GB GPUs. For evaluation, we used multiple sampling and verified functional correctness using the benchmark-provided testbenches executed with Icarus Verilog~\cite{icarus}.

\subsection{Main Results}

Table~\ref{tab:main_results} presents a comprehensive comparison of ReasoningV against existing commercial LLMs, open-source models, and Verilog-specific models across all three benchmarks.

\begin{table*}[htbp]
    \centering
    \begin{threeparttable}
    \caption{Performance comparison across Verilog code generation benchmarks}
    \label{tab:main_results}
    \small
    \begin{tabular}{llcccccccc}
    \toprule
    \multirow{2}{*}{Category} & \multirow{2}{*}{Model} & \multicolumn{3}{c}{VerilogEval-human} & \multicolumn{3}{c}{VerilogEval-machine} & \multicolumn{2}{c}{RTLLM} \\
    \cmidrule(lr){3-5} \cmidrule(lr){6-8} \cmidrule(lr){9-10}
    & & pass@1 & pass@5 & pass@10 & pass@1 & pass@5 & pass@10 & pass@1 & pass@5 \\
    \midrule
    \multirow{4}{*}{Commercial LLM} & GPT-4o-mini & 44.2 & 49.4 & 56.4 & 60.1 & 62.2 & 65.0 & 49.2 & 61.5 \\
    & DeepSeek-R1 & \textbf{81.7} & \textbf{88.1} & \textbf{89.7} & \textbf{81.1} & 85.8 & \textbf{87.4} & \textbf{58.5} & \textbf{66.8} \\
    & DeepSeek-V3 & 70.7 & 77.4 & 78.8 & 77.6 & \textbf{86.2} & \textbf{87.4} & 54.9 & 63.7 \\
    & Gemini-2.0-flash & 59.5 & 67.8 & 70.9 & 74.9 & 79.5 & 81.5 & 46.5 & 56.4\\
    \midrule
    \multirow{3}{*}{Open Source Models} & Qwen2.5-Coder-7B & 34.3 & 46.3 & 50.0 & 61.1 & 75.7 & 78.6 & 31.3 & 45.2 \\
    & Qwen2.5-Coder-14B & \textbf{48.2} & \textbf{59.8} & \textbf{63.5} & \textbf{64.9} & \textbf{77.1} & \textbf{79.7} & \textbf{41.6} & 54.1 \\
    & Qwen2.5-Coder-32B & 47.6 & 59.0 & 61.5 & 61.5 & 64.1 & 75.5 & 40.2 & \textbf{55.7} \\
    \midrule
    % \midrule
    % Open Source Models & Qwen2.5-coder-7B & 34.3 & 46.3 & 50.0 & 61.1 & 75.7 & 78.6 & 31.3 & 45.2 \\
    % \midrule
    \multirow{4}{*}{\parbox{2.5cm}{Verilog-Specific \\ Models}}  & RTLCoder-DeepSeek & 41.6 & 50.1 & 53.4 & 61.2 & 76.5 & 81.8 & 33.5 & 43.9 \\
    & BetterV-CodeQwen & 46.1 & 53.7 & 58.2 & 68.1 & 79.4 & 84.5 & - & - \\
    & OriGen-DeepSeek & \textbf{54.4} & \textbf{60.1} & \textbf{64.2} & \textbf{74.1} & \textbf{82.4} & \textbf{85.7} & - & \textbf{65.5} \\
    & OriGen-Gen-LoRA & 47.4 & 59.9 & 63.4 & 70.2 & 80.2 & 82.7 & \textbf{44.1} & 54.0 \\
    \midrule
    \multirow{2}{*}{RV (Ours)} 
    % & RV-S1 & 51.3 & 60.4 & 63.4 & 64.5 & 78.6 & 81.1 & 42.5 & 55.6 \\
    % & RV-RO & 45.5 & 61.9 & 68.6 & 65.7 & 79.9 & 83.9 & 33.3 & 54.3 \\
    % & RV-Medium & 54.3 & 65.5 & 69.9 & 68.4 & 80.5 & \textcolor{XZW}{83.9} & \textcolor{XZW}{42.6} & \textcolor{XZW}{56.7} \\
    & RV-Adaptive & 53.0 & 64.6 & 67.9 & \textbf{73.6} & 82.8 & \textbf{85.3} & \textbf{45.4} & 56.0 \\
    & RV-Complete & \textbf{57.8} & \textbf{69.3} & \textbf{72.4} & \textbf{73.6} & \textbf{83.4} & \textbf{85.3} & 44.6 & \textbf{62.2} \\
    \bottomrule
    \end{tabular}
    \begin{tablenotes}
    \small
    \item Abbr.: RV = ReasoningV. Bold values indicate best results within each category.
    % \item Abbr.: RV = ReasoningV, S1 = Stage1, RO = Reasoning-Only. Bold values indicate best results within each category.
    \end{tablenotes}
    \end{threeparttable}
\end{table*}

Our best model, ReasoningV-Complete (representing the full two-stage training with the Hard reasoning mode), demonstrates SOTA performance among open-source and Verilog-specific models. It achieves 57.8\% pass@1 on VerilogEval-human, 73.6\% pass@1 on VerilogEval-machine, and 44.6\% pass@1 on RTLLM.

While large commercial models, particularly the reasoning-focused DeepSeek-R1~\cite{deepseekr1}, exhibit superior performance, ReasoningV significantly narrows the gap compared to other models of similar scale (7B parameters). Notably, ReasoningV-Complete outperforms its base Qwen2.5-coder-7B model by substantial margins: +23.5\% on VerilogEval-human, +12.5 pp on VerilogEval-machine, and +13.3\% on RTLLM for pass@1 accuracy. This clearly demonstrates the effectiveness of our dataset curation and training methodology.

Comparing the base open-source models, Qwen2.5-Coder-14B shows better performance than the 7B version, but interestingly, the 32B version does not consistently outperform the 14B model, particularly on VerilogEval-machine and RTLLM pass@5. This suggests that simply increasing model size does not guarantee better Verilog generation performance, highlighting the potential advantage of our parameter-efficient training methods.

The OriGen model~\cite{origen} represents another recent approach to improving Verilog generation through a two-LoRA architecture: Gen LoRA for initial code generation and Fix LoRA for error correction. For fair comparison of pure code generation capabilities (without self-reflection), we evaluated against OriGen using only its Gen LoRA component, which already achieves strong performance with a 47.4\% pass@1 on VerilogEval-human through its code-to-code augmentation methodology. Despite OriGen's impressive results, ReasoningV-Complete still outperforms it by 10.4 percentage points (57.8\% vs. 47.4\%), demonstrating the effectiveness of our approach that combines high-quality verified data (ReasoningV-5K) with full-parameter reasoning enhancement. This performance gap is particularly noteworthy as both models begin with 7B parameter base models but take different approaches to enhancement—ReasoningV's two-stage training with full-parameter optimization versus OriGen's parameter-efficient LoRA adaptation with data augmentation.

\subsection{Ablation Studies}

We conducted comprehensive ablation studies to analyze the impact of different components in the ReasoningV framework, focusing on the effectiveness of the two-stage training methodology and the adaptive reasoning mechanism.

\subsubsection{Effectiveness of Two-Stage Training}

Table~\ref{tab:ablation_training} presents the results of our ablation study on the training methodology, comparing models trained with different stages of our proposed approach against the base model.

\begin{table}[htbp]
    \centering
    \begin{threeparttable}
    \caption{Ablation study on the effectiveness of two-stage training methodology.}
    \label{tab:ablation_training}
    \setlength{\tabcolsep}{2pt}
    \begin{tabular}{lcccccc}
    \toprule
    \multirow{2}{*}{Model} & \multicolumn{2}{c}{VerilogEval-human} & \multicolumn{2}{c}{VerilogEval-machine} & \multicolumn{2}{c}{RTLLM}\\
    \cmidrule(lr){2-3} \cmidrule(lr){4-5} \cmidrule(lr){6-7}
    & pass@1 & pass@5 & pass@1 & pass@5 & pass@1 & pass@5 \\
    \midrule
    Base  & 34.3 & 46.3 & 61.1 & 75.7 & 31.3 & 45.2 \\
    {RV-CT} & 51.3 & 60.4 & 64.5 & 78.6 & 42.5 & 55.6 \\
    {RV-RET} & 45.5 & 61.9 & 65.7 & 79.9 & 33.3 & 54.3 \\
    RV-Complete & \textbf{57.8} & \textbf{69.3} & \textbf{73.6} & \textbf{83.4} & \textbf{44.6} & \textbf{62.2} \\
    \bottomrule
    \end{tabular}
    \begin{tablenotes}
    \small
    \item Abbr.: {CT = Commonsense Training}, {RET = Reasoning Enhancement Training}.
    \end{tablenotes}
    \end{threeparttable}
\end{table}

The results clearly demonstrate the effectiveness and necessity of our two-stage training approach:

\begin{itemize}
\item \textbf{RV-CT (Stage 1 Only)}: Fine-tuning the base model with LoRA on the large OriGen dataset (Stage 1) significantly improves performance across all benchmarks compared to the base model (e.g., +17.0\% pass@1 on VerilogEval-human). This confirms that the first stage effectively establishes foundational knowledge of Verilog syntax and common hardware design principles.

\item \textbf{RV-RET (Stage 2 Only)}: Training the base model directly with full parameters only on the ReasoningV-5K dataset (Stage 2) shows improvement over the base model but surprisingly underperforms compared to RV-CT on VerilogEval-human and RTLLM pass@1. This suggests that while the reasoning-focused dataset is valuable, enhanced reasoning capabilities alone cannot compensate for insufficient foundational domain knowledge.

\item \textbf{RV-Complete (Stage 1 + Stage 2)}: The full two-stage training demonstrates superior performance across all benchmarks, achieving substantial gains over both the base model and the single-stage models (e.g., +6.5\% pass@1 over RV-CT on VerilogEval-human). This confirms that the sequential approach—building foundational knowledge first (Stage 1), followed by targeted reasoning enhancement using the high-quality ReasoningV-5K dataset (Stage 2)—produces optimal results.
\end{itemize}

These findings validate our hypothesis that robust Verilog code generation requires both strong domain knowledge and deep reasoning capabilities, which are best developed through a sequential, staged training approach leveraging appropriate datasets and tuning methods for each stage.

\begin{figure*}[t] % Changed to figure* and adjusted placement (e.g., [t] for top)
    \centering
    \includegraphics[width=\textwidth]{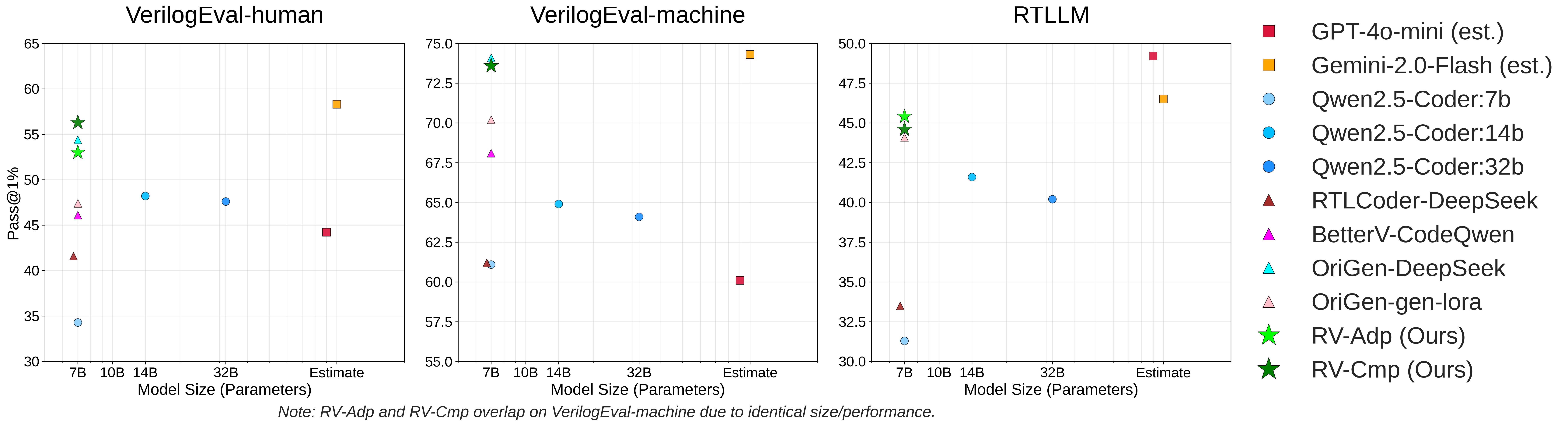}
    \caption{Pass@1 Performance vs. Model Size on VerilogEval-Human.}
    \label{fig:performance_vs_size} % You might want a more specific label if content changed
\vspace{-1em}
\end{figure*}

\subsubsection{Analysis of Adaptive Reasoning Mechanism}

Table~\ref{tab:adaptive_reasoning} presents a comparison of different reasoning approaches applied to the final RV-Complete model. We include fixed difficulty settings (forcing Easy: direct generation, Medium: standard reasoning, or Hard: extended reasoning mode) and our proposed adaptive reasoning mechanism (RV-Adaptive). The table also shows the average number of generated tokens per problem for each approach across the benchmarks. Token counts show averages with percentage changes relative to the adaptive mode. {Additionally, results for DeepSeek-R1 are provided for reference.}

\begin{table}[htbp]
\centering
\caption{Performance and efficiency of reasoning modes on Verilog benchmarks}
\label{tab:adaptive_reasoning}
\footnotesize
\setlength{\tabcolsep}{4pt}
\begin{tabular*}{\hsize}{@{}l 
  c@{\hskip 4pt}c@{\hskip 6pt}  % VerilogEval-H
  c@{\hskip 4pt}c@{\hskip 6pt}  % VerilogEval-M
  c@{\hskip 4pt}c               % RTLLM
@{}}
%\begin{tabular*}{\hsize}{@{}l cc@{\hspace{0.2em}}cc@{\hspace{0.2em}}cc@{}}
\toprule
\multirow{2}{*}{Mode} & \multicolumn{2}{c}{VerilogEval-H} & \multicolumn{2}{c}{VerilogEval-M} & \multicolumn{2}{c}{RTLLM} \\
\cmidrule(r){2-3} \cmidrule(lr){4-5} \cmidrule(l){6-7}
 & P@1 & Tokens & P@1 & Tokens & P@1 & Tokens \\
\midrule
\textit{DS-R1} & 81.7 & 6070 \scriptsize(+584\%) & 81.1 & 3311 \scriptsize(+680\%) & 58.5 & 12686 \scriptsize(+1377\%) \\
\textit{Easy} & 32.1 & \textbf{235} \scriptsize(-74\%) & 51.3 & \textbf{164} \scriptsize(-61\%) & 29.7 & \textbf{356} \scriptsize(-59\%) \\
\textit{Medium} & 54.3 & 1302 \scriptsize(+47\%) & 68.4 & 1156 \scriptsize(+173\%) & 42.8 & 1650 \scriptsize(+92\%) \\
\textit{Hard} & \textbf{57.8} & 2473 \scriptsize(+178\%) & \textbf{73.6} & 1725 \scriptsize(+307\%) & 44.6 & 3940 \scriptsize(+359\%) \\
\textbf{Adaptive} & 53.0 & 888 & \textbf{73.6} & 424 & \textbf{45.4} & 859 \\
\bottomrule
\end{tabular*}

\begin{tablenotes}
\footnotesize
\item \textbf{H}=human, \textbf{M}=machine. P@1 denotes pass@1 metric. \textbf{DS}=DeepSeek.
\end{tablenotes}
\end{table}

The results highlight the significant efficiency benefits of our adaptive reasoning mechanism:
\begin{itemize}
\item \textbf{Token Efficiency}: RV-Adaptive achieves substantial reductions in average token consumption compared to both commercial models and our forced reasoning modes. Compared to DeepSeek-R1, our adaptive approach uses 85\% fewer tokens on VerilogEval-human, 87\% fewer on VerilogEval-machine, and 93\% fewer on RTLLM, while still delivering competitive performance. Even within our own model variants, RV-Adaptive demonstrates remarkable efficiency, using 32\% fewer tokens than Forced Medium and 64\% fewer tokens than Forced Hard across all benchmarks. The savings are even more pronounced on VerilogEval-machine (63\% vs. Forced Medium, 75\% vs. Forced Hard) and RTLLM (48\% vs. Forced Medium, 78\% vs. Forced Hard).
\item \textbf{Performance Trade-off}: RV-Adaptive demonstrates an effective balance between performance and efficiency. On VerilogEval-human, there is a modest pass@1 decrease of 4.8\% compared to always using the Hard mode (Forced Hard), but it still significantly outperforms the base model and RV-CT. Notably, on the more structured VerilogEval-machine benchmark, RV-Adaptive achieves the \textit{same} pass@1 performance as Forced Hard while using 75\% fewer tokens. On RTLLM, RV-Adaptive slightly outperforms Forced Hard in pass@1 (45.4\% vs 44.6\%), potentially due to more appropriate reasoning allocation.
\end{itemize}

Fig.~\ref{fig:performance_vs_size} illustrates the relationship between model size (parameters) and performance (pass@1 on VerilogEval-Human) for the base Qwen2.5-Coder models compared to our ReasoningV variants.

As depicted in the figure, performance generally improves with increasing model size among the base models, although the 32B model shows slightly lower performance than the 14B model on this specific benchmark, suggesting potential diminishing returns from simply increasing scale.

Most notably, the visual comparison highlights that our ReasoningV-Complete model (built upon the 7B parameter base) significantly outperforms the scaling trend observed in the base models. Despite having less than a quarter of the parameters of the 32B base model, ReasoningV-Complete achieves substantially higher performance (57.8\% pass@1 on VerilogEval-Human, as shown in Table~\ref{tab:ablation_training}, compared to the base models' trend). This clearly demonstrates that our targeted approach—leveraging high-quality data curation, the proposed two-stage training methodology, and adaptive reasoning—is a more effective and parameter-efficient strategy for improving Verilog code generation than merely scaling up the model size.

\section{Discussion and Conclusion}

Our experimental results provide strong validation for the ReasoningV framework and its three core innovations. The high-quality ReasoningV-5K dataset, two-stage training methodology, and adaptive reasoning mechanism collectively enable significant performance improvements in Verilog code generation while maintaining computational efficiency.

ReasoningV systematically addressed three critical challenges in hardware design automation: the deficiency of high-quality training data, limited reasoning capabilities for complex hardware design tasks, and computational inefficiency during inference. The large performance gain observed when moving from the base model to ReasoningV-Complete (+23.5\% on VerilogEval-human) highlights the critical impact of our approach. The ablation studies confirm that both training stages contribute substantially to the final performance, with the first stage establishing essential foundational knowledge and the second stage enhancing reasoning capabilities.

Our experimental evaluations demonstrated that ReasoningV substantially advances the SOTA for open-source Verilog generation models. Built on a 7B parameter base model, ReasoningV-Complete achieved 57.8\% pass@1 on VerilogEval-human, 73.6\% pass@1 on VerilogEval-machine, and 44.6\% pass@1 on RTLLM, outperforming previous open-source models by significant margins (+10.4\% over previous best). Notably, ReasoningV-Complete surpassed not only its base model but also much larger models like Qwen2.5-Coder-32B, demonstrating that our targeted approach can be more effective than simply scaling model size.

The adaptive reasoning mechanism offers a practical solution to the computational inefficiency problem, significantly reducing resource requirements (up to 78\% token reduction) while maintaining strong performance. The benchmark-specific adaptive performance suggests that our Judge Adapter effectively identifies when extensive reasoning is necessary versus when a more direct approach suffices. Notably, for the more structured VerilogEval-machine benchmark, adaptive reasoning achieves identical performance to the full Hard mode while using only a quarter of the tokens.

Our error analysis reveals that conceptual understanding of hardware design intent remains the primary challenge for further improvements. Despite these advances, several limitations remain. The ReasoningV-5K dataset, while high-quality, is relatively small and may not cover the full diversity of hardware design challenges. The functional verification employed ensures behavioral correctness but does not guarantee synthesizability or timing compliance. Additionally, there is still a performance gap compared to top commercial models, suggesting room for further improvement.

Future work could explore several promising directions: (1) scaling the ReasoningV-5K dataset while maintaining its quality standards; (2) applying the ReasoningV methodology to larger base models, particularly the 14B parameter models that showed promising scaling characteristics; (3) developing more sophisticated adaptive reasoning mechanisms with finer-grained complexity assessment and budget allocation; (4) integrating feedback from downstream EDA tools into the training and inference process; and (5) extending the approach to support other HDLs and complex design tasks.

ReasoningV represents a significant step toward more reliable, capable, and efficient AI tools for hardware design automation. By addressing the foundational challenges of data quality, reasoning depth, and computational efficiency through targeted methodological innovations, our work provides a pathway for harnessing the potential of LLMs to assist in the increasingly complex task of hardware design. The ReasoningV model, dataset, and associated code are publicly available to foster further research and development in this rapidly evolving field.

\bibliographystyle{IEEEtran}
\bibliography{Main}

\end{document}